\documentclass[aps,prl,reprint,amsfonts,amssymb,amsmath]{revtex4-1}

\usepackage{graphicx}
\usepackage{hyperref}
\usepackage{pdfpages}

\usepackage{color}
\usepackage{bbm}
\usepackage{bm}

\newcommand{\bogband}{\eta} 
\newcommand*{\MyChi}{\raisebox{0.35ex}{\( \chi \)}}%
\newcommand*{\MyChiPP}[1]{\raisebox{0.35ex}{\( \chi^{\prime\prime}_{{#1}} \)}}%
\newcommand{\Ep}{E^{(2,+)}}
\newcommand{\Em}{E^{(2,-)}}
\newcommand{\bz}{b_{z}}

\begin{document}

\title{Topological effects on transition temperatures and response functions in three-dimensional Fermi superfluids}

\author{Brandon M. Anderson}
\email{brandona@uchicago.edu}
\author{Chien-Te Wu}
\author{Rufus Boyack}
\author{K. Levin}
\affiliation{James Franck Institute, University of Chicago, Chicago, Illinois 60637, USA}

\begin{abstract}
We investigate the effects of topological order on 
the transition temperature, $T_c$, and response functions 
in fermionic superfluids with Rashba spin-orbit coupling 
and a transverse Zeeman field in three dimensions.
Our calculations, relevant to the ultracold atomic Fermi gases,
include fluctuations beyond mean-field
theory and are compatible with $f$-sum rules. 
Reminiscent of the $p_x + i p_y$ superfluid, the topological phase is stabilized when driven away
from the Bose-Einstein condensation and towards the BCS limit.
Accordingly,
while experimentally accessible,
$T_c$ is significantly suppressed in a
topological superfluid.
Above $T_c$, the spin and density response functions
provide signatures of topological phases via
the recombination or amplification of frequency dependent
peaks.
\end{abstract}

\maketitle

\textit{Introduction.}$-$
The excitement surrounding topological superfluids~\cite{ReadGreen,KaneRMP,ZhangRMP,Alicea} 
derives from both their scientific as well as 
technological potential. Inspired by the canonical topological superfluid, a spinless $p_{x}+ip_{y}$ superfluid~\cite{ReadGreen},
it has been argued~\cite{Kane,dasSarma1,Sato}
that some combination of spin-orbit coupling (SOC), Zeeman field, as
well as superfluid pairing can artificially produce such a state.
This was explored via the proximity effect~\cite{Kane} in solid state systems  
and using intrinsic pairing in ultracold atomic Fermi gases~\cite{Zhaireview,dasSarma1,Iskin,phasesep1,phasesep2}.
There is, moreover, widespread interest
in experimental confirmations of topological 
band-structure signatures~\cite{Chuanwei2,Chuanwei3,ourtheoryHanPu,DeMelo1,Kitagawa}.

This leads to the goals of this paper. Here we aim to determine
(in the case of intrinsic pairing)
how a transition from a trivial to a topological phase
is reflected in the superfluid transition temperature $T_{c}$.
Additionally, we show how and when a transition in the band-structure can be
experimentally detected via studies of the finite temperature density-density
and spin-spin correlation functions.
The ultracold atomic Fermi gases are ideally suited
for tuning between trivial and topological
phases. As in the
$p_{x}+ip_{y}$ superfluid,
from the perspective of $T_c$,
we find these intrinsically paired superfluids self-consistently adjust to
stabilize topological phases in the BCS regime. This occurs despite the
fact that moderate SOC reinforces
BEC behavior through enhanced pairing~\cite{Fermions1,Fermions2,Fermions3,Fermions4},
even in the normal state~\cite{chientewu,ourtheoryHanPu,ShenoyFluctuations}.

Experiments require the consideration of non-zero temperature $T$.
Although studies of the ground state have been the focus~\cite{Chuanwei2}, 
finite $T$ effects
have been included in the literature~\cite{HanPu,Chuanwei3}
at the mean-field level.
A major weakness of this approach is that
computing $T_c$ in this manner does not reflect 
the topological band-structure, which depends importantly on
the existence of a normal state pairing gap. 
Here we remedy this
inadequacy through the inclusion of 
fluctuations~\cite{ourreview,chientewu},
and also establish that $T_c$ is experimentally accessible.

There are proposals in the literature which suggest that 
the topological phase might
be observed in atomic Fermi gases through the compressibility
$\kappa$~\cite{ourtheoryHanPu,Chuanwei3} 
or via radio frequency (RF) based probes~\cite{Chuanwei2}. 
However, changes in $\kappa$ appear to reflect topology 
only in the limit of small SOC~\cite{ourtheoryHanPu,Chuanwei3}.
RF experiments in principle measure the electronic dispersion,
but resolution and finite temperature broadening effects are not 
yet~\cite{Jin6} well controlled.
Here we introduce an alternative probe: the frequency
dependent density-density or spin-spin correlation functions~\cite{ValePRL08} 
at temperatures $T>T_c$. The position or threshold of peaks
in these responses, importantly, reflects band-structure.   
In the topological phase we find that
a peak in the density response is significantly amplified due to a saddle
point Van Hove singularity, often seen in correlated superfluids~\cite{Kao,Rice}.
In the trivial phase the spin response exhibits two distinct peaks, 
which merge into a single peak in the topological phase.

\textit{Background theory.}$-$
We consider a Fermi superfluid described by the single particle Hamiltonian $H_{0}(\mathbf{k}) =\xi_{\mathbf{k}}+\mathbf{h}\left(\mathbf{k}\right)\cdot\bm{\sigma}$,
where $\xi_{\mathbf{k}}=k^2/2m-\mu$ describes a free particle of momentum $\mathbf{k} = (k_x, k_y, k_z)$, mass $m$, and chemical potential $\mu$. The vector $\mathbf{h}\left(\mathbf{k}\right)=\mathbf{h}_{\bot}\left(\mathbf{k}\right)+\mathbf{h}_{\parallel}\left(\mathbf{k}\right)$ couples the spin-1/2 operator $\bm{\sigma}=({\sigma_x,\sigma_y,\sigma_z})$ to a Zeeman field via $\mathbf{h}_{\parallel}\left(\mathbf{k}\right)=b_{z}\hat{z}$ and an in-plane SOC field 
$\mathbf{h}_{\bot}\left(\mathbf{k}\right)=\lambda \mathbf{k}_{\bot}/m$ for 
in-plane momentum  $\mathbf{k}_\bot = (k_x, k_y, 0)$ and SOC strength $\lambda$. 
Throughout we set $\hbar = k_B = 1$.

The many-body Hamiltonian is of the Bogoliubov-de Gennes (BdG) form
\begin{align}
\mathcal{H}_{\mathrm{BdG}}=\begin{pmatrix}H_{0}\left(\mathbf{k}\right) & \Delta \\
\Delta^{*} & -\widetilde{H}_{0}\left(\mathbf{k}\right)\end{pmatrix},
\label{eq:HBdG}
\end{align}
where $\Delta$ is a pairing gap and 
$\widetilde{H}_{0}\left(\mathbf{k}\right)=\sigma_{y}\left[H_{0}^{*}\left (-\mathbf{k}\right)\right]\sigma_{y}$
is the time-reversed single-particle (hole) Hamiltonian. 
There are four branches in the BdG eigenvalue spectrum, 
$\bogband E_{\alpha {\bf k}}$ for $\alpha, \bogband=\pm1$ with the positive 
energy dispersion
\begin{equation}
E_{\alpha {\bf k}}
=\sqrt{\xi_{\mathbf{k}}^{2}+\left|\mathbf{h}\right|^{2}+
\Delta^{2}+2\alpha\sqrt{\xi_{\mathbf{k}}^{2}\left|\mathbf{h}\right|^{2}+\Delta^{2}b_{z}^{2}}}.
\end{equation}
A three-dimensional superfluid described by the above BdG Hamiltonian belongs to one of three distinct topological phases. 
The topological phase diagram is specified by inequalities derived from solving
$E_{-}(k_z, k_\bot=0)=0$.
No nodes appear when $b_z < \Delta$, corresponding to
a non-topological or ``trivial" superfluid.
If $\mu>0$ and $(\mu^2+\Delta^2)>b_z^2>\Delta^2$, the topological
superfluid has four nodes (4-Weyl points) which
emerge at $k_z^2 = \mu \pm \sqrt{b_z^2-\Delta^2}$.
Finally, for arbitrary $\mu$, the system is a topological superfluid with two nodes
(2-Weyl points) when $b_z^2>(\mu^2+\Delta^2)$~\cite{DeMelo1,HanPu,Chuanwei3}. 
For Rashba SOC, the dispersion around the nodes is linear in momentum, and
is described by a Weyl Hamiltonian with topologically protected nodes.

To compute the transition temperature $T_c$, we build on the
well established mean-field theory~\cite{DeMelo1,HanPu,Chuanwei3}, 
and incorporate fluctuation effects
in a consistent fashion~\cite{Chen2,ourreview}. 
We write the mean-field gap equation~\cite{HanPu,Chuanwei3,Chuanwei2} as
\begin{align}
\Gamma^{-1}\left(0\right)=
&\frac{1}{2}\sum_{\mathbf{k}}\sum_{\bogband\alpha\alpha^{\prime}}
\left(\frac{\delta_{\bogband,+1}-\left(\bogband f\left(E_{\alpha {\bf k}}\right)+f\left(\xi_{\alpha^{\prime} {\bf k}}\right) \right)}
{\bogband E_{\alpha {\bf k}}+\xi_{\alpha^{\prime} {\bf k}}}\right)
\nonumber\\\times& v_{\bogband\alpha\alpha^{\prime}}\left(\mathbf{k},\mathbf{k}\right)+g^{-1}=0,
\label{eq:GM}
\end{align}
where $f(x)$ is a Fermi distribution and $g<0$ is an attractive interaction. Where relevant, we regularize
integrals by introducing a scattering length defined through $g^{-1} = m/4\pi a - \sum_\mathbf{k} m/{\mathbf{k}^2}$~\cite{KetterleReview}.
The coherence factor $v_{\bogband\alpha\alpha^{\prime}}\left(\mathbf{k},\mathbf{k}\right)$
(and its generalization,
$v_{\bogband\alpha\alpha^{\prime}}\left(\mathbf{k},\mathbf {k}-\mathbf{q}\right)$),
is presented in the Supplemental Material~\cite{Supplement}. Their specific
form is irrelevant for the present discussion.

One has to distinguish $T_c$ from the lowest temperature, denoted
$T^*$, at which the mean-field gap equation satisfies
$\Delta(T^*) = 0$.
Such an analysis requires a natural extension~\cite{ourreview,chientewu} 
of Eq.~(\ref{eq:GM}) to finite $Q\equiv(i\omega, \mathbf{q})$
(where $i\omega$ is a Matsubara frequency):
\begin{align}
\Gamma^{-1}\left(Q\right)=&\frac{1}{2}\sum_{\mathbf{k}}\sum_{\bogband\alpha\alpha^
{\prime}}\left(\frac{\delta_{\bogband,+1}-
\left(\bogband f\left(E_{\alpha {\bf k}}\right)+f\left(
\xi_{\alpha^{\prime} \mathbf{k}-\mathbf{q}}\right)
\right)
}{\left(\bogband
E_{\alpha {\bf k}}+\xi_{\alpha^{\prime} \mathbf{k}-\mathbf{q}}\right) - i\omega}\right)
\nonumber \\
\times &v_{\bogband\alpha\alpha^{\prime}}\left(\mathbf{k},\mathbf
{k}-\mathbf{q}\right)+g^{-1}.
\label{eq:GMQ}
\end{align}
From the structure of Eq.~(\ref{eq:GM}) it is apparent that 
$\Gamma\left(0\right)$ depends on both the
full energy spectrum 
$E_{\alpha {\bf k}}$
as well as the bare energy
$\xi_{\alpha {\bf k}}$.
Thus, one might expect (as implemented in Eq.~(\ref{eq:GMQ})),  
that the fluctuation corrections
should depend on an asymmetric combination of
bare and dressed Green's functions
\footnote{However, within the widely used saddle point approximation, the effective vertex obtained via the path integral can only include Green's functions in a symmetric manner.}.

The quantity 
$\Gamma\left(Q\right)$
has been of interest~\cite{HanPu,Fermions1}
for computing the binding energy and mass of the
pairs associated with the isolated two-body physics. 
We emphasize that to describe pairs which are intrinsic to the many-body system,
one should not set 
$\Delta$ and $\mu$ to zero, as is usually done~\cite{HanPu,Fermions1};
the many-body state is
not simply a gas constructed from entities of the
two-body problem.
In 3D there are metastable
or resonant pairs 
for all parameters below $T^{*}$, whereas the two-body
bound states only exist for positive scattering length.
Furthermore,
for sufficiently large $b_{z}$,
the effective mass in the two-body problem diverges near
unitarity~\cite{HanPu}, and is not defined at negative scattering length.

To characterize the mass of the many-body
system, consider the vertex
in Eq.~(\ref{eq:GMQ}) expanded at small momenta, where (using Eq.~(\ref{eq:GM}))
$
\Gamma\left(Q\right)\approx a_{0}^{-1}(i\omega-\omega_{\mathbf{q}})^{-1},
$
with $a_{0}=\left(\partial_{i\omega}\Gamma\left(Q\right)\right|_{Q=0}$.
The pair dispersion is
$\omega_{\mathbf{q}}=a_{0}^{-1}\left(\Gamma\left(0,\mathbf{q}\right)-\Gamma\left(0,0\right)\right)
\approx q_{\perp}^2/2M_\perp + q_{\parallel}^2/2M_{\parallel}$,
where $M_{\perp}$ ($M_{\parallel}$) is the effective pair mass for the
component of momentum parallel (perpendicular) to the SOC vector. 
While it is sometimes possible to calculate the effective pair masses
$M_{\perp}$, $M_{\parallel}$ analytically,
in general this is not necessary. Rather, it suffices to
calculate numerically the second-order derivative
at small $\mathbf{q}$
\footnote{Since $\Gamma\left(0,\mathbf{q}\right)$ will be symmetric in $\mathbf{q}$ for inversion-symmetric systems, the first and third order terms in $\mathbf{q}$ will vanish, and the lowest correction will come in at fourth order.}.

Notice that the vertex function $\Gamma\left(Q\right)$ reflects, up to
a constant, the non-interacting Green's function of a thermal Bose
gas with pair dispersion $\omega_{\mathbf{q}}$, 
below the condensation temperature
($\Gamma^{-1}\left(0\right) = 0$).
We interpret this $ Q \neq 0$ contribution by considering
the following quantity, proportional to a Bose occupation number $n_B$: 
\begin{equation} 
\sum_{Q \neq 0}\Gamma\left(Q\right)
= \Delta^2 = a_0^{-1} n_B,\quad \textrm{at $T=T_c$}.
\label{eq:DLT}
\end{equation}
This characterizes the excitation gap in the limit
in which all pairs are non-condensed, but in which $T_c$ is approached from below.
The Thouless condition (describing the instability of the
normal state) or divergence of $\Gamma(0)$ requires
that the above expression for $\Delta^2$ lies on the mean-field
curve determined from Eq.~(\ref{eq:GM}). 
The condition for $T_c$ is then simply 
obtained~\cite{Chen2,ourreview,chientewu}
by equating the constraint on $\Delta(T_c)$
via Eq.~(\ref{eq:DLT})
with that obtained from the mean-field gap equation in Eq.~(\ref{eq:GM}).

Importantly, all these arguments can be generalized so
that the computation of $T_c$ as a fluctuation correction to \textit{any}
BCS-BEC mean-field theory (including LOFF-like phases)
is now accessible, based on writing
$\Gamma(Q)$ as a natural extension of the appropriate mean-field vertex function
(e.g., Eq.~(\ref{eq:GMQ})).
This approach is distinguished from
other BCS-BEC theories~\cite{NSR,Strinati2,SadeMelo} by the fact that
the transition temperature is dependent on a normal
state excitation gap; in 
this way $T_c$ will reflect changes 
in band-structure associated with the transition from trivial
to topological phases. It is also distinguished by the fact
that the present approach avoids the unphysical first order
transition found in all other BCS-BEC 
theories~\cite{firstordertransitionpapers}.
We will assume throughout that, above $T_c$, the mean-field gap represents a reasonable 
approximation
\footnote{One can improve on this approach, [see J. Maly et al, Physica C 321, 113 (1999) and He et al, PRB 76, 224516 (2007)] but the added complexity does not affect $T_c$ and does not lead to new physics.}
for the normal state $\Delta$.

\begin{figure}
\includegraphics[width=3.3in,clip]
{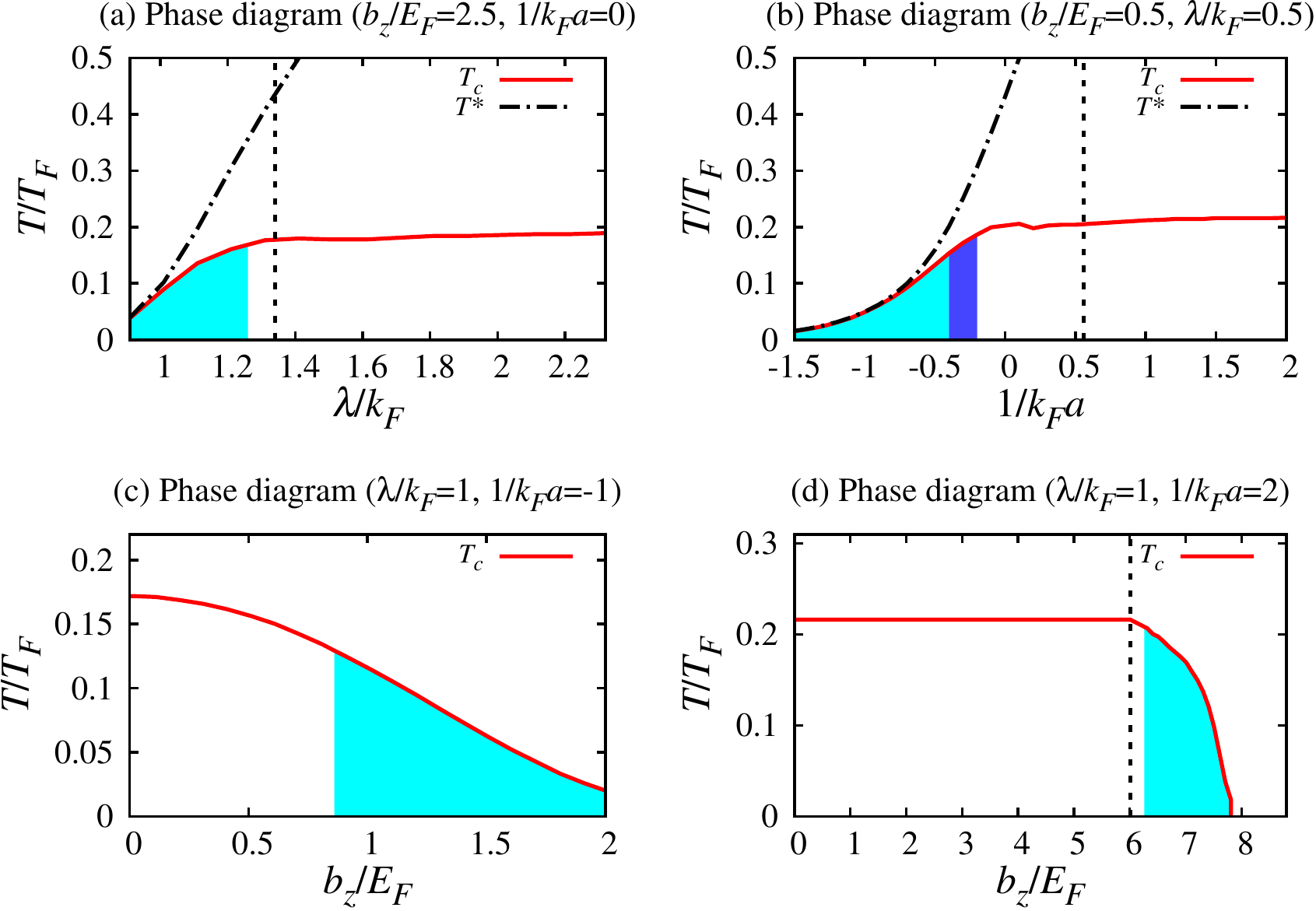}
\caption{
Phase diagrams for the superfluid temperature $T_c$ and (where shown) the mean-field transition temperature $T^*$.
The $T=0$ topological phases are indicated by
shaded regions in light (dark) blue color with
2 (4)-Weyl points. The top two panels show the dependence on either
SOC strength $\lambda$ (left) or scattering length
$1/{k_Fa}$ (right), with other parameters fixed.
The lower panels show $T_c$ vs. $b_z$ with $\lambda/k_F=1$ for both panels, $1/{k_F a}=-1$
on the left and $1/{k_F a}=2$ on the right. Dotted lines indicate
$\delta \mu =0$. Once a topological phase is entered the system becomes more BCS-like.}
\label{fig:PhaseFixBz}
\end{figure}

\textit{Phase diagram.}$-$
To understand the effects of
SOC and the Zeeman field on condensation and pairing,
we numerically compute $T^*$ and $T_c$, 
varying $1/k_F a$, $b_z$, and $\lambda$.
Where relevant, we measure quantities in terms
of the Fermi momentum ($k_F$), energy ($E_F$), or temperature ($T_F$).
It is convenient to define a shifted chemical potential, 
$\delta \mu  = \mu - \mu_0$, 
where $\mu_0 = - \mathrm{max} \left\{ E_{\rm so} (1 + b_{z}^2 / E_{\rm so}^2) / 2, b_{z} \right\}$ 
and $E_{\rm so} = \lambda^2/m$ is the SOC energy. 
In this way, a necessary (but not sufficient) condition for
a topological phase is that
$\delta\mu>0$.

Figure~\ref{fig:PhaseFixBz} plots $T_c$ and (in some cases) the pairing
onset temperature $T^*$ as a function of either $\lambda$, $b_z$, or
$1/k_F a$. Dotted lines indicate where $\delta \mu =0$. Where relevant, these
plots are consistent with earlier work~\cite{ourtheoryHanPu}. A close
analogy between varying $\lambda$ and varying $1/k_Fa$ is seen in
Fig.~\ref{fig:PhaseFixBz}(a) and Fig.~\ref{fig:PhaseFixBz}(b). 
We define ``weak" or ``enhanced" pairing relative to $b_z=0$. The former is associated with
small $\lambda$ or negative $1/k_Fa$ while the latter corresponds to either large $\lambda$
or large positive $1/k_Fa$.
Thus, Fig.~\ref{fig:PhaseFixBz}(c) is characteristic of the generic weak
pairing regime while Fig.~\ref{fig:PhaseFixBz}(d) is characteristic of
the strong pairing case produced by either large $1/k_Fa$ or large
$\lambda$.

We analyze the top two figures by focusing on a
decreasing abscissa which effects a transition from a trivial to
topological phase (shown as shaded). 
In Fig.~\ref{fig:PhaseFixBz}(a),
corresponding to $ 1/k_F a = 0$, $b_z = 2.5 E_F$,
this transition is driven by varying the SOC strength 
$\lambda$. In
Fig.~\ref{fig:PhaseFixBz}(b)
it is driven directly by varying the scattering length $1/k_{F}a$; 
somewhat after
the point
$\delta \mu > 0$ is crossed, a further \textit{decrease} in $1/k_Fa$ (towards the
BCS limit) allows the
system to reach a topological phase.
Here we see a series of two transitions from topologically
trivial to 4-Weyl and then to 2-Weyl superfluids.
While there is some initial decline in $T_c$ with
diminishing $1/k_Fa$, 
the most significant decrease in $T_c$ occurs in the 2-Weyl case.

The next two panels 
contrast the regime of weak pairing 
(Fig.~\ref{fig:PhaseFixBz}(c))
with that of enhanced pairing
(Fig.~\ref{fig:PhaseFixBz}(d)).
In the first case, the system is BCS-like everywhere. Increasing $b_z$ gradually suppresses $T_c$
and there is no clear signature in $T_c$ of the change from a trivial to 
a topological phase (shown as shaded in the figure).
As shown in Fig.~\ref{fig:PhaseFixBz}(d),
when the pairing is enhanced,
$T_c$ becomes insensitive to
variations in the Zeeman field until $\delta \mu = 0$.
Shortly thereafter, the topological phase transition is crossed and 
$T_c$ rapidly declines.

We can see from the last figure, in particular,
that the satisfaction of the topological inequality and the $\delta \mu = 0$
condition importantly define a transition 
(often quite sharp, as in
Fig.~\ref{fig:PhaseFixBz}(d))
between a superfluid, characterized by a larger gap, and larger pair mass,
$M_{\perp} \sim 2 m$ (i.e., more ``BEC-like"), and a superfluid with a small gap, $\Delta/E
_F \ll 1$, and a small pair mass $M_{\perp} \ll m$ which is ``BCS-like".
The resulting behavior of $T_c$ arises
in the topological phase because there is a competition
between the effects of a decreasing pair mass
and a decreasing
mean-field pairing gap as $b_z$ increases. The net effect is a lowering
of $T_c$ in the topological phase. This can, in turn, be viewed as a form
of BEC-BCS transition.
The details are presented in the Supplemental Material~\cite{Supplement}.

One can inquire as to why the topological transition becomes more apparent 
(as reflected in $T_c$) 
on the strong pairing side
(Fig.~\ref{fig:PhaseFixBz}(d)),
whereas it is less evident (from the perspective
of $T_c$) when in the weak pairing limit
(Fig.~\ref{fig:PhaseFixBz}(c)).
These differences are
reflected in the evolution of the
band-structure via a Van Hove singularity
as the topological transition is crossed.
To address this, Fig.~2 presents a constant energy contour plot for
the band $+E_{-1, {\bf k}}$.
The two axes correspond to the in-plane 
($k_\bot$) and out-of-plane ($k_z$) momenta. 
For definiteness, we have chosen
$1/k_Fa = 0$ and $\mu(T)$, $\Delta(T)$ are determined for a temperature just above $T_c$. 
Local extrema in this figure reflect Van Hove singularities,
either at isolated points or extended in a ring-like structure.
Each of the three panels in a given row
corresponds to increasing values of
$b_z$ with only the left-most figures in the trivial phase.
The top three figures are in the weak pairing
regime whereas the bottom
three figures are in the regime of enhanced pairing.

A key observation from these figures is that in the weak pairing limit
there is a smooth
evolution from a trivial to topological phase, 
whereas for enhanced pairing
the band-structure
evolves rather dramatically from a trivial and BEC-like phase
to a topological and BCS-like phase. 
Indeed, the topological transition
in the lower panel is roughly correlated with the appearance of additional
Van Hove singularities (as indicated).
This is in contrast to the upper panel where Van Hove singularities 
of the trivial and topological phases are relatively unchanged. 
These figures help interpret the behavior observed in Fig.~1(c)
and Fig.~1(d).

\textit{Frequency dependent spin and density response
functions.}$-$
As in previous work~\cite{chientewu} we write
the correlation functions (above $T_c$) as
\begin{align}
\MyChi_{S_{i}S_{j}}\left(i \omega,\mathbf{q}\right)=&\sum_{\mathbf{k}}\sum_{\alpha\alpha',\bogband\bogband'}\left(\frac{f(\bogband E_{\alpha\mathbf{k}})-f(\bogband' E_{\alpha'\mathbf{k+q}})}{\bogband E_{\alpha\mathbf{k}}-\bogband'E_{\alpha'\mathbf{k+q}}+i\omega}\right)\nonumber\\&\times w_{\alpha\alpha',\bogband\bogband'}(\mathbf{k},\mathbf{k+q}).
\label{eq:CHPP}
\end{align}
The density-density correlation function $\MyChi_{\rho\rho}(Q)$ corresponds to $i=j=0$, with $\sigma_{0}=\mathbbm{1}_{2},$
whereas $i,j\in\{x,y,z\}$ gives the corresponding spin-spin correlation function.
The differences between the density or spin responses are the coherence factors
$w_{\alpha\alpha',\bogband\bogband'}(\mathbf{k},\mathbf{k+q})$,
which are rather complicated and are presented in the Supplemental Material. 
As a numerical check on these calculations, the $f$-sum rule
for the density response and related sum rules \cite{chientewu}
for the spin response hold for all $\mathbf{q}$.

\begin{figure}
\includegraphics[width=3.3in,clip]{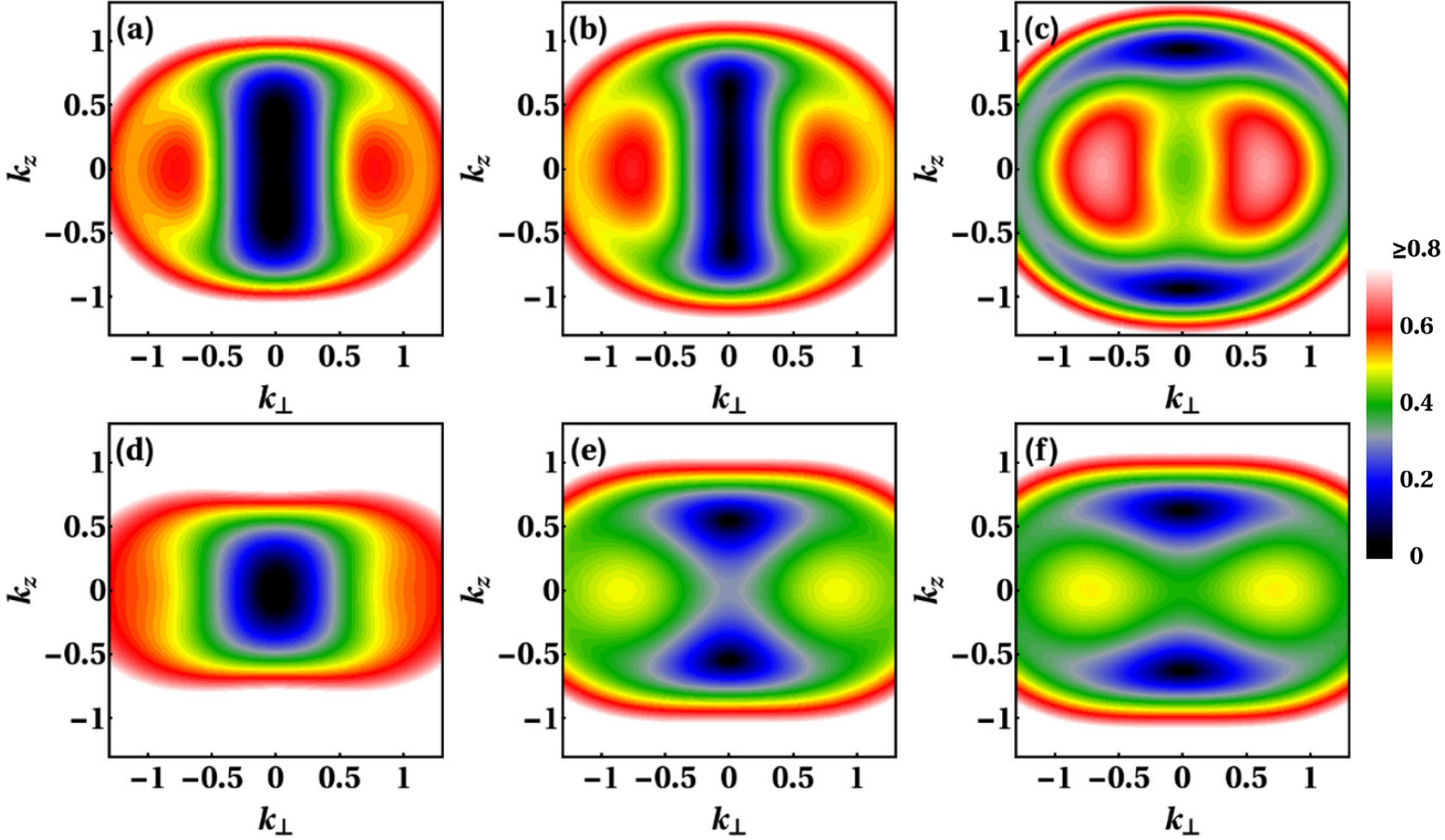}
\caption{
Evolution of the dispersion as the topological transition is crossed by tuning $\bz$. 
In the weak pairing limit (top panel), the system smoothly evolves across the transition, whereas
for enhanced pairing (bottom panel) there is a more abrupt change in band-structure.
In all plots we show constant energy contours $+E_{-1,\bf k}/E_F$ at unitarity, with
$k_\perp$ and $k_z$ in units of $k_F$.
For panels (a)-(c) we set $\lambda/k_F=0.5$ and the Zeeman field
$b_z/E_F=0.4, 0.6, 0.8$, whereas for panels (d)-(f) we set $\lambda/k_F=1$ and
$b_z/E_F=1.2, 1.7, 1.8$ respectively. Only the left-most figures are in a trivial phase.}
\label{fig:contour}
\end{figure}

Quite generally, the correlation functions for a paired
normal state can be decomposed into two parts; one involving the
difference: $\Em(\mathbf{k,q})=|E_{-1,\mathbf{k}}-E_{\pm1,\mathbf{k+q}}|$
which enters as a thermal contribution (at $T \neq 0$), and the other
involving the sum: $\Ep(\mathbf{k,q})= |E_{-1,\mathbf{k}}+E_{\pm 1,\mathbf{k+q}}|$,
which we call the multiparticle contribution.
We address the $\mathbf{q}={0}$ spin response,
$\MyChi_{S_{i}S_{j}}(\omega,0)$, (where $i, j$ are $x$ or $y$)
so that inter-band terms dominate.
Thus, for the $\pm 1$ subscript in the density response, the $-1$ band
label yields the main contribution,
whereas in the spin response the $+1$ band label is most important.

Figure 3(a) shows $\MyChi_{S_{x}S_{y}}(\omega,0)$ 
for both the trivial and topological phases. 
In the trivial phase there are two clearly resolvable peaks; the first peak is associated with
the thermal contribution and the second with the multiparticle contribution.
By contrast, there is only
one peak in the topological phase. 
A related signature 
for the Hall conductivity (in 2D) at
$T=0$, rather than, as here, above $T_c$, 
was suggested earlier~\cite{Kitagawa}.

Importantly, this provides
a means of distinguishing between the trivial and
topological phases.
We can analytically identify the position of the maximum
in the first (thermal) peak, which is due to a flat band in $\Em(\mathbf{k},0)$, and appears at precisely $2b_z$.
The threshold for the second peak is $\omega_{1}\equiv\mathrm{min}_{\mathbf{k}}
\Ep(\mathbf{k},0).$ In the trivial phase we find that, if $\mu>0$, $\omega_{1}=2\Delta$, 
whereas if $\mu<0$, $\omega_{1}=2(\Delta^2+\mu^2)^{1/2}.$ 
Hence $\omega_{1}$
is strictly greater than the frequency of the first peak (2$b_{z}$), 
thus yielding two distinct peaks in the response function.
In the topological phase,
$\omega_{1}=2b_z$ so that the two
peaks merge.

\begin{figure}
\includegraphics[width=3.3in,clip]{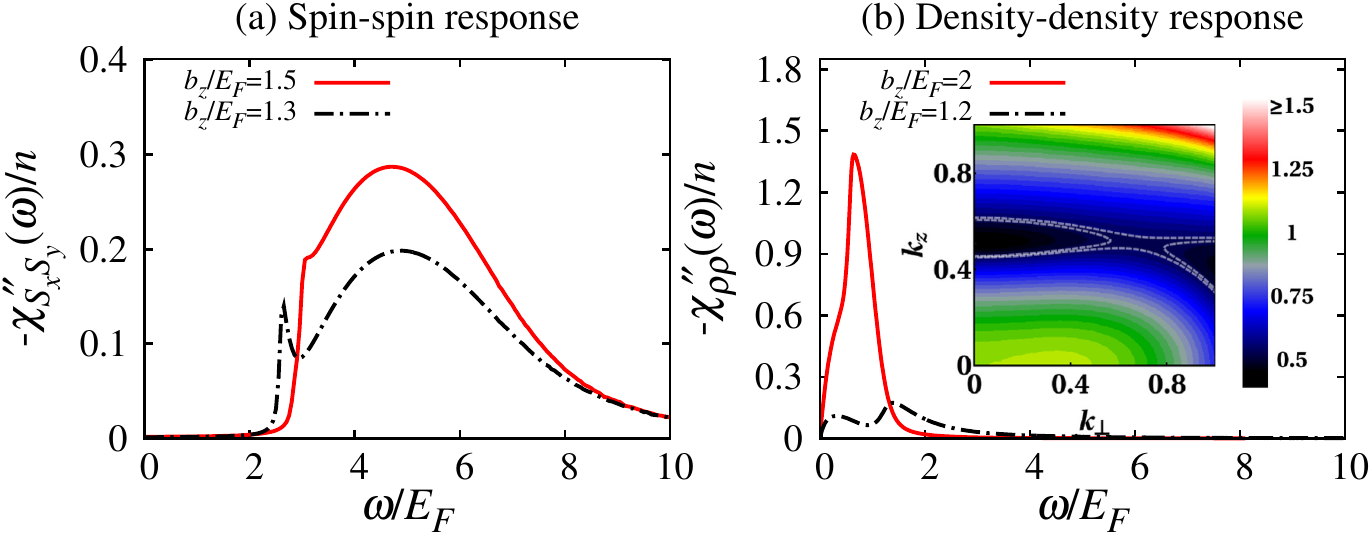}
\caption{Contrast between topological (solid, red) and trivial (dashed, black) phases
of the frequency dependent spin-spin (Fig.~\ref{fig:chi}(a)) and density-density (Fig.~\ref{fig:chi}(b)) correlation functions.
Both response functions are calculated at $1/k_Fa=0$ and $\lambda/k_F = 1$, with respective wave-vectors of 
$\mathbf{q}={0}$ and $\mathbf{q}=0.5k_F\hat{z}$ for the spin and charge responses.
The inset in Fig.~\ref{fig:chi}(b) shows the energy contours of $\Ep(\mathbf{k},\mathbf{q})/E_F$ in the topological phase, with $k_\perp$ and $k_z$ in units of $k_F$.
The dashed lines highlight the saddle point Van Hove
singularity whose magnitude determines the frequency location of the peak response in Fig.~\ref{fig:chi}(b).}
\label{fig:chi}
\end{figure}

We now focus on the density-density correlation
function $\MyChi_{\rho\rho}(\omega,\mathbf{q})$, which is only non-zero when $\mathbf{q}\neq0$.
This is shown in Fig.~3(b) for the case of unitarity: $1/k_Fa = 0$,
and we can again compare the trivial and topological phases.
Here $\lambda/k_F=1$ and we
plot the imaginary part of the response function, $\MyChiPP{\rho\rho}(\mathbf{q},\omega)$ \cite{chientewu, Supplement},
deep in the topological phase ($b_z/E_F=2$)
and in the trivial phase ($b_z/E_F=1.2$) at $\mathbf{q}=0.5k_F\hat{z}$ and $T=0.21 T_F$ (just above $T_c$).

In the trivial phase there are two peaks, one associated with
thermal contributions involving
$\Em(\mathbf{k,q})$
and the second with the multiparticle
component involving 
$\Ep(\mathbf{k,q})$.
In the topological phase, there is a large
peak at $\omega/E_{F} = 0.6$, which arises from
a (2D) saddle point Van Hove singularity contribution
in $\Ep(\mathbf{k},\mathbf{q})$. This is associated with
$\nabla_{\mathbf{k}} \Ep(\mathbf{k},\mathbf{q})=0$,
which (via the density of states)
enters as a denominator in the response functions.
These saddle point Van Hove singularity effects are well known~\cite{Kao,Rice}
and are illustrated in the inset on the right and discussed in the
Supplemental Material.
Importantly, here we observe that as the system enters the
topological phase they amplify the peaks in the density-density
correlation function, thus helping to distinguish 
between the trivial and topological phases. 

\textit{Conclusions.$-$}
This paper addresses how an intrinsically produced
condensation temperature varies across a topological
transition, induced by varying SOC, Zeeman coupling, or the scattering length.
Importantly, the introduction of fluctuations necessarily introduces a feedback
of the topological band-structure into $T_c$. 
The passage from the trivial to the topological phase is accompanied by
a transition in which the system is driven towards a low $T_c$,
more
BCS-like phase with smaller pair mass and smaller gap. 
Nevertheless, there is a range of $\bz$ in the topological phase where $T_c\sim 0.1T_F$,
which is experimentally accessible~\cite{KetterleReview}.

We also present methods of detecting the topological band-structure above $T_c$,
exploiting frequency dependent peaks in the density and spin responses. 
The topological transition appears in the spin response as a recombination
of two peaks, which are separate in the trivial phase. In the topological
superfluid, the dynamical density response exhibits
a greatly amplified peak associated with a (2D) saddle point Van Hove singularity.
In both the response functions and $T_c$ we find
that the topological transition appears
quite smooth in the weak pairing  
and much more abrupt and apparent in the strong pairing regime.

\textit{Acknowledgements.}$-$
This work was supported by NSF-DMR-MRSEC 1420709.

\textit{Note added.}$-$
Recently, we became aware of a complementary paper that considers fluctuation effects in spin-orbit coupled superfluids with fixed relative population density using a closely related formalism~\cite{HeFluctuations}.

\bibliography{Review}

\clearpage


\section*{Supplementary material}

# Supplemental Material: Topological effects on transition temperatures and response functions in three-dimensional Fermi superfluids

Brandon M. Anderson, Chien-Te Wu, Rufus Boyack, and K. Levin
*James Franck Institute, University of Chicago, Chicago, Illinois 60637, USA*

## I. DERIVATION OF THE VERTEX FUNCTION $\Gamma(Q)$

Here we present derivations of the vertex function $\Gamma(Q)$, where $Q \equiv (i\omega, \mathbf{q})$, along with the coherence factor $v_{\eta\alpha\alpha'}(\mathbf{k}, \mathbf{k} - \mathbf{q})$, which appear in Eq. (4) of the main text. We begin by writing the non-interacting Green's function in terms of projectors as follows

$$G_0(K) = \sum_\alpha \frac{P_\alpha^0(\mathbf{k})}{i\nu - \xi_{\alpha\mathbf{k}}}, \tag{S1}$$

where $K \equiv (i\nu, \mathbf{k})$ and $P_\alpha^0(\mathbf{k}) = \frac{1}{2} U_\mathbf{k}(1 + \alpha\sigma_z) U_\mathbf{k}^\dagger$ is a projector into the single-particle band $\alpha = \pm 1$. The matrix $U_\mathbf{k}$ is the unitary operator that diagonalizes $H_0(\mathbf{k})$ to produce the single-particle dispersion $\xi_{\alpha\mathbf{k}} = \xi_\mathbf{k} + \alpha|\mathbf{h}|$.

Similarly, the Nambu Green's function $\mathcal{G}(K)$ for a superfluid can be written in terms of projectors as

$$\mathcal{G}(K) = [i\nu - \mathcal{H}_{\text{BdG}}(\mathbf{k})]^{-1} = \begin{pmatrix} G(K) & F(K) \\ \widetilde{F}(K) & \widetilde{G}(K) \end{pmatrix} = \sum_{\eta\alpha} \frac{\mathcal{P}_{\eta\alpha}(\mathbf{k})}{i\nu - \eta E_{\alpha\mathbf{k}}}, \tag{S2}$$

where we have used the inverse of the BdG Hamiltonian, $\mathcal{H}_{\text{BdG}}$, to define the normal and anomalous Green's functions $G(K)$ and $F(K)$, along with their time-reversed counterparts $\widetilde{G}(K) = i\sigma_y [G(-K)]^T i\sigma_y$ and $\widetilde{F}(K) = \sigma_y [F(-K)]^T \sigma_y$. The projectors $\mathcal{P}_{\eta\alpha} = \psi_{\eta\alpha}\psi_{\eta\alpha}^\dagger$ are constructed from the BdG eigenvectors

$$\psi_{\eta\alpha} = \mathcal{U}_\mathbf{k} \begin{pmatrix} \alpha\sqrt{\frac{1}{2}\left(1 + \alpha\frac{\xi_\mathbf{k}}{E_{0\mathbf{k}}}\right)} \sqrt{\frac{1}{2}\left(1 + \alpha\eta\frac{\zeta_{\alpha\mathbf{k}}}{E_{\alpha\mathbf{k}}}\right)} \\ \eta\sqrt{\frac{1}{2}\left(1 - \alpha\frac{\xi_\mathbf{k}}{E_{0\mathbf{k}}}\right)} \sqrt{\frac{1}{2}\left(1 - \alpha\eta\frac{\zeta_{\alpha\mathbf{k}}}{E_{\alpha\mathbf{k}}}\right)} \\ \alpha\eta\sqrt{\frac{1}{2}\left(1 + \alpha\frac{\xi_\mathbf{k}}{E_{0\mathbf{k}}}\right)} \sqrt{\frac{1}{2}\left(1 - \alpha\eta\frac{\zeta_{\alpha\mathbf{k}}}{E_{\alpha\mathbf{k}}}\right)} \\ \sqrt{\frac{1}{2}\left(1 - \alpha\frac{\xi_\mathbf{k}}{E_{0\mathbf{k}}}\right)} \sqrt{\frac{1}{2}\left(1 + \alpha\eta\frac{\zeta_{\alpha\mathbf{k}}}{E_{\alpha\mathbf{k}}}\right)} \end{pmatrix}, \tag{S3}$$

where $\mathcal{U}_\mathbf{k} = \text{diag}\{U_\mathbf{k}, V_\mathbf{k}\}$ rotates the particle (hole) sector to the spin-orbit basis with a unitary matrix $U_\mathbf{k}$ ($V_\mathbf{k}$), and we have defined $\theta = \cos^{-1}(b_z/|\mathbf{h}|)$, $E_{0\mathbf{k}} = \sqrt{\xi_\mathbf{k}^2 + \Delta^2 \cos^2\theta}$, and $\zeta_{\alpha\mathbf{k}} = E_{0\mathbf{k}} + \alpha|\mathbf{h}|$. Note that, up to a sign, $\zeta_{\alpha\mathbf{k}}$ limits to $\xi_{\alpha\mathbf{k}}$ as $\Delta \to 0$ or $b_z \to 0$, and to $E_{\alpha\mathbf{k}}$ as $\lambda \to 0$.

For convenience, the $4 \times 4$ projector matrices can be expressed as four $2 \times 2$ sub-matrices as

$$\mathcal{P}_{\eta\alpha}(\mathbf{k}) \equiv \begin{pmatrix} P_{\eta\alpha}(\mathbf{k}) & Q_{\eta\alpha}(\mathbf{k}) \\ R_{\eta\alpha}(\mathbf{k}) & S_{\eta\alpha}(\mathbf{k}) \end{pmatrix}. \tag{S4}$$

The Green's function $G(K)$ is found from the appropriate $2 \times 2$ sub-matrix with the corresponding projector

$$P_{\eta\alpha}(\mathbf{k}) = \frac{1}{4E_{0\mathbf{k}}E_{\alpha\mathbf{k}}} U_\mathbf{k} \begin{pmatrix} (E_{0\mathbf{k}} + \alpha\xi_\mathbf{k})(E_{\alpha\mathbf{k}} + \alpha\eta\zeta_{\alpha\mathbf{k}}) & \alpha\Delta^2 \sin\theta\cos\theta \\ \alpha\Delta^2 \sin\theta\cos\theta & (E_{0\mathbf{k}} - \alpha\xi_\mathbf{k})(E_{\alpha\mathbf{k}} - \alpha\eta\zeta_{\alpha\mathbf{k}}) \end{pmatrix} U_\mathbf{k}^\dagger. \tag{S5}$$

We can now define a quantity $\chi(Q)$, known as the pair susceptibility, which has been introduced in previous papers [S1]; it is a natural extension of $\chi(0)$ and appears in $\Gamma(Q)$:

$$\chi(Q) \equiv -\frac{1}{2} \text{Tr}\left[\sum_K G(K) \widetilde{G}_0(K - Q)\right], \tag{S6}$$

where $\widetilde{G}_0(K) = i\sigma_y [G_0(-K)]^T i\sigma_y$ is the time-reversed, or hole, non-interacting Green's function.

Substituting the above definitions then gives

$$\chi(Q) = -\frac{1}{2}\mathrm{Tr}\left[\sum_K \sum_{\eta\alpha} \frac{P_{\eta\alpha}(\mathbf{k})}{i\nu - \eta E_{\alpha\mathbf{k}}} \sum_{\alpha'} \frac{P^0_{\alpha'}(\mathbf{k}-\mathbf{q})}{i\nu - i\omega + \xi_{\alpha'\mathbf{k}-\mathbf{q}}}\right],$$

$$= -\frac{1}{2}\sum_{\mathbf{k}}\sum_{\eta\alpha\alpha'}\left(\sum_{i\nu}\frac{1}{i\nu - \eta E_{\alpha\mathbf{k}}}\frac{1}{i\nu - (i\omega - \xi_{\alpha'\mathbf{k}-\mathbf{q}})}\right)\mathrm{Tr}\left[P_{\eta\alpha}(\mathbf{k})P^0_{\alpha'}(\mathbf{k}-\mathbf{q})\right]. \quad (S7)$$

Performing the summation over Matsubara frequencies reduces this expression to

$$\chi(Q) = \frac{1}{2}\sum_{\mathbf{k}}\sum_{\eta\alpha\alpha'}\left(\frac{\delta_{\eta,+1} - (\eta f(E_{\alpha\mathbf{k}}) + f(\xi_{\alpha'\mathbf{k}-\mathbf{q}}))}{(\eta E_{\alpha\mathbf{k}} + \xi_{\alpha'\mathbf{k}-\mathbf{q}}) - i\omega}\right)v_{\eta\alpha\alpha'}(\mathbf{k},\mathbf{k}-\mathbf{q}). \quad (S8)$$

Here the coherence factor which appears in Eq. (3) and Eq. (4) of the main text is

$$v_{\eta\alpha\alpha'}(\mathbf{k},\mathbf{k}-\mathbf{q}) = \mathrm{Tr}\left[P_{\eta\alpha}(\mathbf{k})P^0_{\alpha'}(\mathbf{k}-\mathbf{q})\right]. \quad (S9)$$

The vertex function can now be defined by $\Gamma(Q) \equiv \left[\chi(Q) + g^{-1}\right]^{-1}$. Using the expression for the susceptibility in Eq. (S8), we obtain the vertex function $\Gamma(Q)$ as given in Eq. (4) of the main text. The familiar gap equation can then also be obtained from the Thouless criterion: $\chi(0) + g^{-1} = 0$.

## II. THE TOPOLOGICAL PHASE DIAGRAM IN THE ENHANCED PAIRING LIMIT

Figure S1 shows the major findings concerning the topological phase diagram in the enhanced pairing limit. In Fig. S1(a) the transition temperature $T_c$ and the pairing onset or mean-field transition temperature $T^*$ are plotted as functions of the Zeeman field $b_z$. Here we consider the limit of enhanced pairing through large SOC strength $\lambda$, which leads to a more BEC-like behavior at $b_z = 0$. (We could also have considered small $\lambda$ with sufficiently large $1/k_F a$ without qualitatively affecting the result.) In Fig. S1(b) the pair mass, which is an important component of $T_c$, is plotted as a function of $b_z$. Finally, in Fig. S1(c) we present a plot of the zero temperature mean-field gap as a function of this Zeeman field. Notable here is that for a large range of $b_z$, the pair mass $M_\perp(T=0)$, [S2] the mean-field gap $\Delta(T=0)$, and the transition temperature $T_c$, are only weakly dependent on $b_z$. Nevertheless, Fig. S1(a) shows a striking contrast between the two temperatures $T_c$ and $T^*$. Whereas $T^*$ decreases continuously with Zeeman field, $T_c$ tends to be relatively constant until the point $\delta\mu = 0$ is crossed, after which point the topological phase transition is encountered and $T_c$ starts to rapidly decline. This demonstrates that $T^*$ is completely independent of topology, whereas $T_c$ clearly reflects the topological phase transition.

Now we give insight into the behavior of $T_c$, which depends on the solutions to the mean-field gap and number equations and on the behavior of the pair mass. From the perspective of $T_c$, we find that the trivial to topological transition reflects a BEC-BCS transition. This is reminiscent of other topological superfluids, such as a $p_x + ip_y$

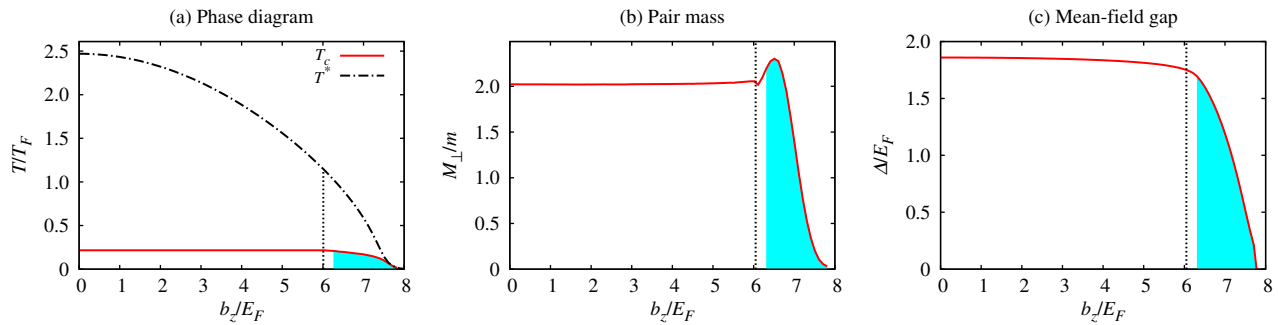

FIG. S1. Plots of the (a) $T_c$ and $T^*$ phase diagram, (b) zero temperature pair mass, $M_\perp(T=0)$, and (c) zero temperature mean-field gap, $\Delta(T=0)$, as a function of $b_z$ in the enhanced pairing limit ($1/k_F a = 2$ with $\lambda/k_F = 1$). In all figures the topological superfluid phase is shaded blue, and a dotted line separates the region with $\delta\mu < 0$ (left) and $\delta\mu > 0$ (right). The superfluid transition temperature $T_c$ is insensitive to $b_z$ until $\delta\mu > 0$. However, the mean-field transition temperature $T^*$ monotonically decreases with increasing $b_z$. The weak dependence of $T_c$ on $b_z$ is correlated to similar behavior of the pair mass shown in (b) and the mean-field gap at zero temperature shown in (c).

superfluid, where the topologically non-trivial phase occurs on the BCS side of the BCS-BEC crossover. Ultimately we will show that the central physics arises from the behavior of $\Gamma(Q)$ and $\Gamma(0)$, which, through the coherence factors implicitly contain the BdG wave-function. In this way the change in topology is correlated to changes in $T_c$.

### A. Dependence of the transition temperature $T_c$ on the Zeeman field $b_z$

Our aim now is to understand how $T_c$ depends on $b_z$. While mean-field theory would predict a phase transition at $T^*$, $T_c$ is crucially dependent on the pair mass. To make this discussion more concrete, we refer to Fig. S2 which indicates how the transition temperature is deduced. The procedure is as follows. First we define the quantity $\Delta_{\text{pg}}^2(T)$,

$$\Delta_{\text{pg}}^2(T) = \sum_{Q \neq 0} \Gamma(Q), \quad T \leq T_c, \tag{S10}$$

which is commonly referred to as the pseudogap. From Eq. (5) in the main text, we see that $T_c$ is the point at which this function coincides with the mean-field gap $\Delta^2(T)$. Figure S2(a) presents a plot of the mean-field gap vs. temperature for two different values of $b_z$ which are associated with the topological phase. For these two values of $b_z$, Fig. S1 shows that both the pair mass and the transition temperature are decreasing rather rapidly with increasing $b_z$. The dotted lines which intersect these curves are plots of $\Delta_{\text{pg}}^2(T)$ and the points of intersection represent $T_c$ for each value of $b_z$. Also labeled is $T^*$ for each value of $b_z$. It is also clear that the slope of the lines is greater for the lower $b_z$ case. This is because the pair mass is larger. Consequently the fluctuation suppression of $T_c$ relative to $T^*$ is, thereby, greater. This leads to a larger separation between $T^*$ and $T_c$ at lower $b_z$. Nevertheless, because $T^*$ is so much larger here, $T_c$ is as well. Indeed, Fig. S2(a) shows that even in the narrow range of $b_z$, the transition temperature rapidly drops, as observed in Fig. S1(a).

Figure S2(b) is a similar representation of our transition temperature calculations for two values of $b_z$ in the trivial phase. This regime is associated with the BEC limit where one can see that both the pair mass and $T_c$ are very nearly independent of Zeeman field. This should be contrasted with the behavior of $T^*$ for these two values of $b_z$, which shows a substantial decline with increasing $b_z$.

Of the four values of $b_z$ illustrated in Fig. S2(a) and Fig. S2(b), the higher $b_z = 7.6 E_F$ is clearly in a BCS-like limit where the pair mass is small and $T_c \approx T^*$. The lower two values of $b_z$ ($b_z/E_F = 2, 5$) are in the BEC regime, where $T_c$ and the pair mass are relatively constant. Finally, the case $b_z = 7.3 E_F$ is in an intermediate regime between BCS and BEC-like behavior.

We can understand these BCS and BEC endpoints semi-quantitatively. From the definition of $\Gamma(Q)$ in the text it is straightforward to show that $a_0 \Delta_{\text{pg}}^2(T) = (\bar{M} T/2\pi)^{3/2} \zeta(3/2)$, where $\bar{M} = \left(M_\parallel M_\perp^2\right)^{1/3}$ is the geometric mean of the pair mass and $\zeta(z)$ is the Riemann-zeta function. In the extreme BEC limit (where $\Delta(T_c) \approx \Delta(0)$, $\bar{M}(T_c) \approx \bar{M}(0)$), it follows that $T_c \sim \Delta^2(0)/\bar{M}(0)$, or that the transition temperature scales with the inverse pair mass. This is to be associated with the trivial phase. In the extreme BCS limit we have $\Delta^2(T_c) = c(1 - T_c/T^*)\Delta^2(0)$ (with dimensionless constant $c$). This yields $(1 - T_c/T^*) = (T_c/T^*)^{3/2} \Delta_{\text{pg}}^2(T^*)/c\Delta^2(0)$. Solving for $T_c$ in this small pseudogap (BCS) limit implies that $T_c \sim T^*$, as expected. This describes the situation within the topological phase. Note that this BCS expression is independent of both the pair mass and the gap.

### B. General structure of pair mass

The previous figures indicate that the pair mass is an important component of the present theory. The pair mass can be written as

$$\frac{1}{2M_{ij}} = \frac{1}{2a_0} \frac{\partial^2}{\partial q_i \partial q_j} \sum_{\mathbf{k}} \sum_{\eta \alpha \alpha'} \left( \frac{\delta_{\eta,+1} - (\eta f(E_{\alpha \mathbf{k}}) + f(\xi_{\alpha' \mathbf{k} - \mathbf{q}}))}{\eta E_{\alpha \mathbf{k}} + \xi_{\alpha' \mathbf{k} - \mathbf{q}}} \right) v_{\eta \alpha \alpha'}(\mathbf{k}, \mathbf{k} - \mathbf{q}) \Bigg|_{\mathbf{q}=0}, \tag{S11}$$

where $v_{\eta \alpha \alpha'}(\mathbf{k}, \mathbf{k} - \mathbf{q})$ is the coherence factor defined earlier. By symmetry, the off diagonal elements ($i \neq j$) of $1/M_{ij}$ vanish, and furthermore $1/M_{xx} = 1/M_{yy} \equiv 1/M_\perp$, and we also define $1/M_{zz} \equiv 1/M_\parallel$.

The most significant features of the pair mass, as shown in Fig. S1(b), can be understood from this equation. We start by considering the pole structure. Of the eight denominators of the form $\eta E_{\alpha \mathbf{k}} + \xi_{\alpha' \mathbf{k} - \mathbf{q}}$ at $\mathbf{q} \to 0$, only the term with $\alpha = \alpha' = \eta = -1$ can be zero. The critical value, $b_c$, for which this denominator can vanish is found to be $\mu < 0$ with $b_c = \frac{1}{2}\left(\sqrt{\mu^2 + \Delta^2} + |\mu|\right)$. Note that this condition occurs half-way between the topological phase transition and the $\delta \mu = 0$ transition.

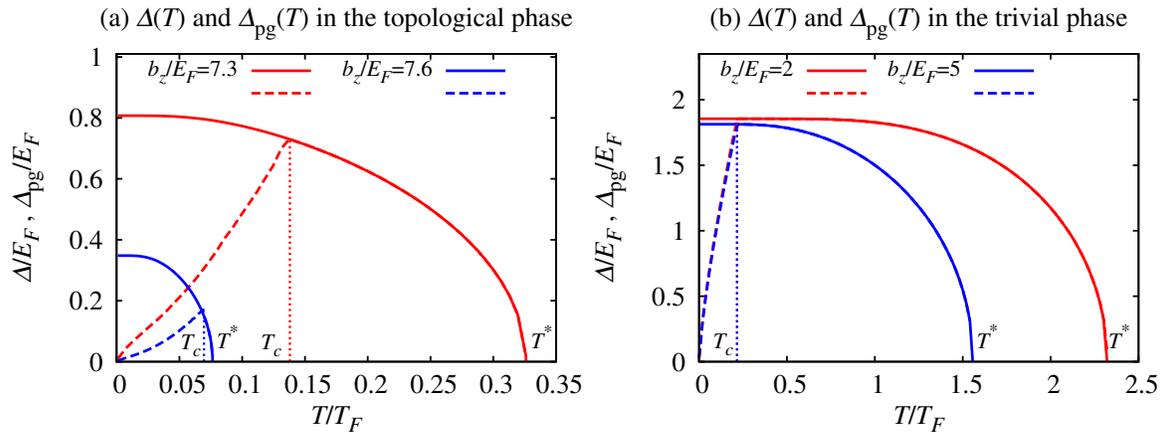

FIG. S2. Dependence of the mean-field gap $\Delta(T)$ (solid curves) and pseudogap $\Delta_{\rm pg}(T)$ (dashed curves) as a function of temperature for the strong coupling limit ($1/k_F a = 2$ and $\lambda/k_F = 1$). The left (right) panel shows the topolgical (trivial) phase respectively. The strength of Zeeman coupling $b_z/E_F$ is indicated in the legend, and a dotted line of the same color corresponds to the projection of $T_c$ where $\Delta(T_c) = \Delta_{\rm pg}(T_c)$. Recall that by definition, $T = T^*$ defines where the mean-field gap vanishes ($\Delta(T^*) = 0$). In the left panel, the stronger Zeeman coupling, $b_z/E_F = 7.6$, leads to a smaller $T^*$. However, the fluctuation effect at $b_z/E_F = 7.6$ is reduced and $T_c$ is pushed toward $T^*$. In other words, $T_c \approx T^*$. In the right panel, as $b_z$ is increased $T^*$ drops significantly whereas $T_c$ does not change. In this case the $\Delta_{\rm pg}$ and $T_c$ curves almost overlap each other for both $b_z$ values. In fact, for this enhanced pairing limit, the pseudogap and $T_c$ curve would be nearly identical for any $b_z < b_c$. Note further that $\Delta(0) \approx \Delta(T_c)$ in this limit.

For $b_z < b_c$, the system is topologically trivial. This regime can be understood if we note that no denominators in the inverse pair mass can be zero. Therefore, the integral is dominated by the coherence factors at small momentum, which will be approximately constant. Furthermore, as we will show, in this same limit the mean-field solution is also strongly insensitive to $b_z$. This combines to produce the constant curve for the topologically trivial phase as observed in the pair mass, and consequently $T_c$.

Shortly after the topological phase transition is crossed, $b_z > b_c$, a line of singularities in the integrand emerges defined implicitly by the curve $\left(b_z^2 - \Delta^2/4\right) = \xi_{\mathbf{k}} |\mathbf{h}(\mathbf{k})|$. After performing the $\mathbf{q}$ derivatives in Eq. (S11), a term will appear with the pole $(-E_{-,\mathbf{k}} + \xi_{-,\mathbf{k}})^{-3}$. At larger $b_z$, this third order singularity will dominate all other terms in the integrand, and the integral is well approximated by only this contribution. This will result in a rapid growth of $1/M_{ij}$ at large $b_z$, which leads to the suppression in the pair mass that was observed numerically.

The crossover between these two regimes is characterized by a small increase in the pair mass. This can be further understood by noting that the lower-order poles contribute with a negative sign. This reduces the inverse pair mass for a small region before the third order pole begins to dominate. Thus, there is a small increase in the pair mass after the condition $b_z = b_c$ is crossed.

## C. Dependence of the mean-field equations on the topological phase transition

The mean-field behavior of the gap function is also an important ingredient in these calculations of $T_c$. We now demonstrate the sensitivity of the mean-field solutions on the condition for the topological phase transition. This should be contrasted with the behavior of $T^*$, which is defined as smallest $T$ that solves the gap equation in the $\Delta = 0$ limit:

$$\sum_{\mathbf{k}}\sum_{\alpha} \frac{(1 - 2f(|\xi_{\mathbf{k}\alpha}|))}{4|\xi_{\mathbf{k}\alpha}|}\left[1 + \alpha\frac{h^2}{|\xi_{\mathbf{k}}|\sqrt{h^2 + (\lambda k_\perp)^2}}\right] = -\frac{1}{g}. \tag{S12}$$

Since $\Delta$ does not enter into the above equation, the topological condition is not relevant for $T^*$; this is in contrast to the behavior of $T_c$.

We now address the mean-field gap ($I_G$) and number ($I_N$) equations and limit our discussion here mainly to the ground state. This is largely for simplicity and because the topological condition is most directly reflected here. Moreover, it is relatively straightforward to generalize to finite $T$. We begin by defining functions such that the self consistent mean-field equations are obtained when $I_G[\Delta(b_z), \mu(b_z), b_z] = I_N[\Delta(b_z), \mu(b_z), b_z] = 0$. Differentiating

these equations with respect to $b_z$, and applying the chain rule, one can find the change in $\Delta$ and $\mu$ with $b_z$ is

$$\begin{pmatrix} \frac{d\Delta}{db_z} \\ \frac{d\mu}{db_z} \end{pmatrix} = - \begin{pmatrix} \frac{\partial I_G[\Delta,\mu,b_z]}{\partial \Delta} & \frac{\partial I_G[\Delta,\mu,b_z]}{\partial \mu} \\ \frac{\partial I_N[\Delta,\mu,b_z]}{\partial \Delta} & \frac{\partial I_N[\Delta,\mu,b_z]}{\partial \mu} \end{pmatrix}^{-1} \begin{pmatrix} \frac{\partial I_G[\Delta,\mu,b_z]}{\partial b_z} \\ \frac{\partial I_N[\Delta,\mu,b_z]}{\partial b_z} \end{pmatrix}. \tag{S13}$$

We can now understand the dependence of the mean-field gap and chemical potential on the Zeeman field $b_z$. Specifically, at zero temperature and in the enhanced pairing limit, *the gap and chemical potential are approximately independent of the strength of the Zeeman field until we cross into a topologically non-trivial phase.* After this point, the gap begins to rapidly drop with Zeeman field, whereas the shifted chemical potential ($\delta\mu$) increases. This demonstrates that these changes can be associated with a BEC to BCS transition.

Analyzing this behavior is most tractable in the limit that $\lambda \to 0$. We consider $I_G$, but a similar argument applies for $I_N$. In this limit, the gap equation is

$$I_G\left[\Delta\left(b_z\right),\mu\left(b_z\right),b_z\right] = \frac{1}{2}\sum_\mathbf{k} \frac{1}{\sqrt{\xi_\mathbf{k}^2 + \Delta^2}} \theta\left(\sqrt{\xi_\mathbf{k}^2 + \Delta^2} - b_z\right) + \frac{1}{g}, \tag{S14}$$

where $\theta(x) = 1$ for $x \geq 0$ and $\theta(x) = 0$ for $x < 0$ is the unit step function. These expressions show that the system is completely insensitive to $b_z$, provided that $\min_\mathbf{k}\left\{\sqrt{\xi_\mathbf{k}^2 + \Delta^2}\right\} \geq b_z$. This condition is precisely the condition for a topologically trivial superfluid. As $b_z$ is increased, the system eventually crosses into a topologically non-trivial phase. In this phase, the step function vanishes for some range of $\mathbf{k}$, allowing the gap equation to be satisfied with smaller $\Delta$. Thus, the system becomes more BCS-like with increasing $b_z$.

If we consider the limit with $\lambda > 0$, we see that similar arguments hold. For example, the coherence factor in $I_G$ contains, in part, a term of the form

$$\left[\sqrt{(\xi_\mathbf{k}^2 + \Delta^2) + (\lambda k_\perp \xi_\mathbf{k}/b_z)^2} + \alpha b_z\right] \xrightarrow{k_\perp \to 0} \left[\sqrt{\xi_\mathbf{k}^2 + \Delta^2} + \alpha b_z\right]. \tag{S15}$$

The $\alpha = -1$ contribution can switch sign for some parameters. Specifically, the minimum condition for this term to change sign occurs at $k_\perp = 0$. Taking this limit gives a condition that is only satisfied in a topologically non-trivial phase.

In addition to the change in sign once the topological condition is crossed, if $\lambda \neq 0$, there is an additional variation with $b_z$ that is not due to the suppression of the integrand in $I_G$ and $I_N$. However, this small shift is not responsible for the more qualitative effect of an evolution from BEC-like to BCS-like physics as the topological phase transition is crossed.

### III. RESPONSE FUNCTIONS IN THE NORMAL PHASE

In this section we derive the density-density and spin-spin correlation functions in the presence of SOC and a Zeeman field. Here we focus on the normal phase ($T > T_c$), which does not require collective-mode contributions. The density-density or spin-spin correlation functions, as in the main text, can be written as $\chi_{S_i S_j}(Q) \equiv \int d\tau\, e^{i\omega\tau} \langle T_\tau S_{\mathbf{q}i}(\tau) S_{-\mathbf{q}j}(0)\rangle$, for a many-body density or spin operator $S_{\mathbf{q}i} = \sum_{ss'\mathbf{k}} c^\dagger_{\mathbf{k}s}(\sigma_i)_{ss'} c_{\mathbf{k}+\mathbf{q}s'}$. Here $i = j = 0$, with $\sigma_0 = \mathbb{1}_2$, corresponds to the density-density correlation function $\chi_{\rho\rho}(Q)$, whereas $i, j \in \{x, y, z\}$ gives the corresponding spin-spin correlation function.

We emphasize that in the normal state there is no anomalous Green's function component, but the existence of normal state pairs allows one to write the correlation functions as the sum of two terms

$$\chi_{S_i S_j}(Q) = \sum_K \text{Tr}\left[\sigma_i G(K) \sigma_j G(K+Q) + \sigma_i F(K) \sigma_j \widetilde{F}(K+Q)\right]. \tag{S16}$$

Here we associate $F(K) = (\Delta/\Delta^*)\widetilde{F}(K)$ with a pseudogap vertex contribution, which leads to

$$\chi_{S_i S_j}(Q) = \sum_K \sum_{\alpha\alpha',\eta\eta'} w_{\alpha\alpha',\eta\eta'}(\mathbf{k}, \mathbf{k}+\mathbf{q})(i\nu - \eta E_{\alpha \mathbf{k}})^{-1}(i\nu + i\omega - \eta' E_{\alpha'\mathbf{k}+\mathbf{q}})^{-1}, \tag{S17}$$

where we have introduced the coherence factor

$$w_{\alpha\alpha',\eta\eta'}(\mathbf{k}, \mathbf{k}+\mathbf{q}) = \text{Tr}[\sigma_i P_{\eta\alpha}(\mathbf{k})\sigma_j P_{\eta'\alpha'}(\mathbf{k}+\mathbf{q})] + \text{Tr}[\sigma_i Q_{\eta\alpha}(\mathbf{k})\sigma_j R_{\eta'\alpha'}(\mathbf{k}+\mathbf{q})]. \tag{S18}$$

Upon performing the summation over Matsubara frequencies, we obtain the expression for the density-density or spin-spin correlation function given in Eq. (6) of the main text:

$$\chi_{S_iS_j}(Q) = \sum_{\mathbf{k}} \sum_{\alpha\alpha',\eta\eta'} \left( \frac{f(\eta E_{\alpha\mathbf{k}}) - f(\eta' E_{\alpha'\mathbf{k+q}})}{\eta E_{\alpha\mathbf{k}} - \eta' E_{\alpha'\mathbf{k+q}} + i\omega} \right) w_{\alpha\alpha',\eta\eta'}(\mathbf{k}, \mathbf{k+q}). \tag{S19}$$

In any theory of correlation functions, it is important to ensure they satisfy the appropriate sum rules, which may or may not arise due to a conservation law. For the density response, the $f$-sum rule is a consequence of charge conservation, which holds even in the presence of SOC and a Zeeman field. The explicit form of the $f$-sum rule is

$$\int_{-\infty}^{\infty} \frac{d\omega}{\pi} \left( -\omega \chi''_{\rho\rho}(\omega, \mathbf{q}) \right) = \frac{nq^2}{m}. \tag{S20}$$

Here, and below, we analytically continue the response to real frequency $\chi_{S_iS_j}(\omega, \mathbf{q}) = \lim_{\delta \to 0} \chi_{S_iS_j}(i\omega, \mathbf{q})|_{i\omega = \omega + i\delta}$, and then define the dissipative part $\chi''_{S_iS_j}(\omega, \mathbf{q})$ of the correlation function. Note that for $i = j$, the dissipative part is equivalent to the imaginary part of the response function, but for $i \neq j$ it may be complex in general [S1]. As can be shown [S1], using charge conservation in the form of the Ward-Takahashi identity, the above density-density correlation function manifestly satisfies the $f$-sum rule.

The associated sum-rule for the $\chi_{S_xS_y}(\omega, \mathbf{q})$ correlation function can be shown [S1] to be

$$\int_{-\infty}^{\infty} \frac{d\omega}{\pi} (-\omega \chi''_{SxSy}(\omega, \mathbf{q})) = 0. \tag{S21}$$

Since $\chi''_{SxSy}(\omega, \mathbf{q})$ is even in frequency, this sum rule is trivially satisfied. An additional sum rule can be derived, however, which is given by a different moment of the spin-spin correlation function; this sum rule is

$$\int_{-\infty}^{\infty} \frac{d\omega}{\pi} \chi''_{SxSy}(\omega, \mathbf{q}) = \langle [S_{\mathbf{q}x}, S_{-\mathbf{q}y}] \rangle = 2i \langle \sigma_z \rangle, \tag{S22}$$

where $\langle \sigma_z \rangle = \sum_K \mathrm{Tr}\, [\sigma_z G(K)]$ is the spin polarization in the $\hat{z}$ direction.

In all our numerical calculations the spin sum rule is satisfied to within one percent and similarly the $f$-sum rule for density is found to be accurate to within five percent.

### A. Van Hove singularities and indications of the topological phase

In the main text we have noted that these two-body probes are to be contrasted with the more frequently studied (one-body) spectral functions, as measured in photoemission or radio frequency experiments. We consider peaks in the response functions as a function of the frequency $\omega$ for a fixed wave-vector $\mathbf{q}$. These are associated with thermal contributions involving $E^{(2,-)}(\mathbf{k}, \mathbf{q})$ and multiparticle contributions involving $E^{(2,+)}(\mathbf{k}, \mathbf{q})$, where we define

$$\begin{aligned} E^{(2,-)}(\mathbf{k}, \mathbf{q}) &= |E_{-1,\mathbf{k}} - E_{\pm 1,\mathbf{k+q}}|, \\ E^{(2,+)}(\mathbf{k}, \mathbf{q}) &= |E_{-1,\mathbf{k}} + E_{\pm 1,\mathbf{k+q}}|. \end{aligned} \tag{S23}$$

We find that the $-1$ sign is most relevant for the density response and the $+1$ sign is most relevant for the $\mathbf{q} = 0$ spin response.

Van Hove singularities lead to important enhancements in the response functions because the density of states enters as a denominator. These singularities correspond to

$$\nabla_{\mathbf{k}} E^{(2,\pm)}(\mathbf{k}, \mathbf{q}) = 0, \tag{S24}$$

and they are associated with minima, maxima or saddle points. Importantly, for a system with rotational symmetry, $k_\perp$ and $k_z$ form an effectively two-dimensional integrand. In such systems, it is the saddle point Van Hove singularity which leads to the most prominent peaks. We can now understand features of Fig. 3(b) in the main text. The enhanced peak of the topological phase is associated with a saddle point Van Hove singularity. The deeper one goes into the topological regime, the more pronounced the saddle point effects are.

Figure S3 presents a contour plot of these constant energy surfaces for $E^{(2,+)}(\mathbf{k}, \mathbf{q})$ which enters into the density response. The different panels indicate how the system evolves from the trivial to the topological phase (left to right). The top (a)-(c) and bottom (d)-(f) panels correspond to weak and enhanced pairing, respectively. The contours

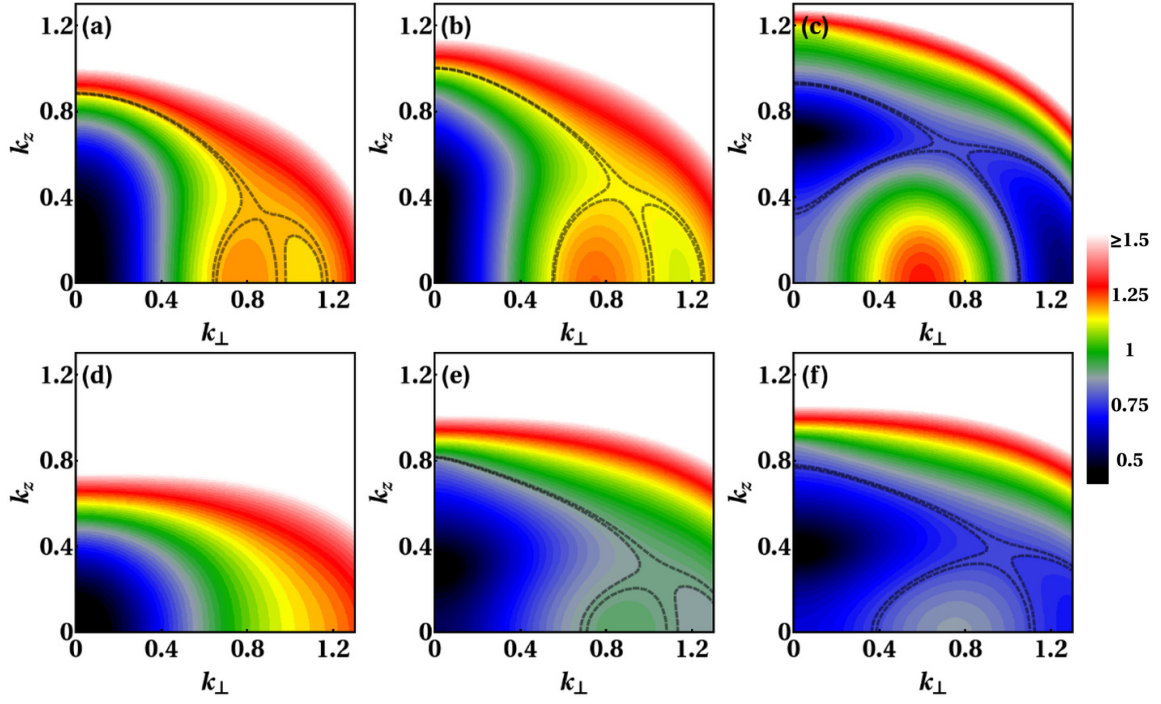

FIG. S3. Evolution of Van Hove singularities (VHS) with $b_z$ in the contours of constant multiparticle energy $E^{(2,+)}(\mathbf{k}, \mathbf{q})$. In the upper panel, $\lambda/k_F = 0.5$, and from left to right, the Zeeman field takes values $b_z/E_F = 0.4$, 0.6, and 0.8, respectively. In the lower panel $\lambda/k_F = 1$, and from left to right $b_z/E_F = 1.2$, 1.7, and 1.8, respectively. In both cases, the leftmost panel is topologically trivial, and the other two are nontrivial. Saddle point VHS are enclosed by the black dashed curves. As evidenced in these figures, for $\lambda/k_F = 0.5$, VHS can exist even before the topological phase is entered while VHS can only appear deep in the topological phase for $\lambda/k_F = 1$. In all figures, $\mathbf{q} = 0.5 k_F \hat{z}$, $1/k_F a = 0$, and $T = 0.21 T_F$, and the wavevectors and multiparticle energy are measured in Fermi units.

denoted by dashed lines enclose the saddle point positions. As in Fig. 2 of the main text, here we observe that in the enhanced pairing regime the onset of a saddle point is closely associated with the trivial to topological phase transition. In either case, once the system is deeper into the topological phase the saddle point Van Hove singularities are enhanced. This, in turn, leads to the peak enhancement in the density-density correlation function shown in Fig. 3(b) of the main text.



\end{document}